\documentclass[letterpaper, 10 pt, conference]{ieeeconf}  

\IEEEoverridecommandlockouts                              
\usepackage{amsmath}
\usepackage{amssymb}
\usepackage{float}
\usepackage{graphicx}
\usepackage{subfigure}
\usepackage{epstopdf}
\usepackage{color}
\usepackage{verbatim}
\usepackage{tikz,times}
\usepackage{algorithm}
\usepackage{algorithmic}
\usepackage{cite}

\DeclareMathOperator{\diag}{diag}

\newcommand{\norm}[1]{\left\lVert#1\right\rVert}

\newtheorem{remark}{\textbf{Remark}}
\newtheorem{theorem}{\textbf{Theorem}}
\newtheorem{proposition}{\textbf{Proposition}}
\newtheorem{lemma}{\textbf{Lemma}}

\newtheorem{definition}{\textbf{Definition}}
\newtheorem{assumption}{\textbf{Assumption}}

\title{\LARGE \bf
Secure distributed filtering for unstable dynamics under compromised observations
}

\author{Xingkang~He,  
	Xiaoqiang Ren,
	Henrik Sandberg,
	Karl Henrik Johansson 
\thanks{The work is supported by Knut \& Alice Wallenberg foundation, and by Swedish Research Council.}
\thanks{X. He, X. Ren, H. Sandberg and K. H. Johansson are with Division of Decision and Control Systems, School of Electrical Engineering and Computer Science. KTH Royal Institute of Technology, Sweden ((xingkang,xiaoqren,hsan,kallej)@kth.se).}%
}

\begin{document}

\maketitle
\thispagestyle{empty}
\pagestyle{empty}

\begin{abstract}

In this paper, we consider a secure distributed filtering problem for linear time-invariant systems with bounded noises and unstable dynamics under compromised observations.
A malicious attacker is able to compromise a subset of the agents and manipulate the observations arbitrarily.
We first propose a recursive distributed filter consisting of two parts at each time. The first part employs a saturation-like scheme, which gives a small gain if the innovation is too large. The second part is a consensus operation of state estimates among neighboring agents.
A sufficient condition is then established for the boundedness of  estimation error, which is with respect to network topology, system structure, and the maximal  compromised agent subset. We further provide an equivalent statement, which connects to 2s-sparse observability in the centralized framework in certain scenarios, such that the sufficient condition is feasible.
Numerical simulations are finally provided to illustrate the developed results.

\end{abstract}

\section{Introduction}
Cyber-physical systems (CPSs) are systems controlled and monitored by computer-based algorithms. Through a CPS,  physical processes and cyber components can be effectively integrated. During the recent years, numerous applications of CPSs such as sensor network, vehicle network, process control, smart grid, etc, have been well investigated in academia and industry.  With higher integration of large-scale computer networks and complex physical processes, the CPSs 
are confronting more security issues both in software and physical layers. 
Thus, the research topics on CPS security  are attracting more and more attention.

In a CPS, sensor observations can be utilized to obtain state estimate or to design output feedback signal to control the physical process. Due to the vulnerabilities of  sensors, the attacker may insert faulty data into observations of the compromised sensors. Then, the estimates or controller based on the compromised observations will be unreliable, and even bring tremendous damage to the whole system. 
Thus, some detection and identification schemes are considered to find out whether the sensors are under attack, and if so how to identify the attack signals  inserted to the systems.
A  study on attack detection and identification for CPSs was given in \cite{pasqualetti2013attack}, where the design methods and analysis techniques for centralized and distributed monitors were discussed as well.
In \cite{forti2018distributed}, the joint distributed attack detection and state estimation was investigated in a Bayesian framework.
To obtain attack-resilient state estimates, in the centralized framework,  some state estimators or observers were proposed based on optimization techniques \cite{pajic2017attack,pajic2017design,fawzi2014secure,shoukry2017secure,han2019convex}, 
recursive implementation \cite{nakahira2018attack}, and probabilistic approach \cite{ren2019secure}.
Compared with centralized methods, on one hand, the distributed ones have advantages in quite a few aspects, such as the structure robustness, energy saving and parallel processing. 
On the other hand, in the distributed framework, since each agent has limited information from local observations and neighboring communications, the distributed state estimation methods are essentially different from the centralized ones. 
In the distributed state estimation under compromised sensors, observer-based methods were studied for byzantine attacks, under which the compromised sensors can send faulty information to other normal sensors \cite{mitra2016secure}. 
In \cite{deghat2019detection}, a distributed observer with attack detection layer was proposed to deal with a class of biasing attacks.
Distributed estimation for a static parameter under compromised observations was studied in \cite{chen2018topology}, where the sparse-observability condition was required to guarantee the consistency of the estimator.

In this paper, we study the secure distributed filtering or estimation problem for linear time-invariant systems with bounded noises and unstable dynamics.
The main contributions of this paper are three-fold.
1) We investigate the secure distributed filtering problem under compromised observations.  Unlike \cite{deghat2019detection}, we allow that the malicious attacker manipulates the observations arbitrarily for  an unknown subset of the agents.
Different from \cite{mitra2016secure,mitra2018secure} requiring some robustness of communication graph,  we simply assume the connectivity of the graph.
2) We propose a novel secure distributed filtering framework consisting of two parts, which is essentially different from the centralized methods \cite{pajic2017attack,pajic2017design,fawzi2014secure,shoukry2017secure,han2019convex,ren2019secure,nakahira2018attack} or the distributed methods \cite{deghat2019detection,mitra2016secure}. The first part employs a saturation-like scheme, which gives a small gain if the innovation is too large. The second part is a consensus operation of state estimates among neighboring agents. 
3) Different from the analysis methods for static parameter estimation \cite{chen2018topology,chen2018resilient}, we provide a new  analysis approach for unstable system dynamics, and establish a sufficient condition for the boundedness of  estimation error. We further provide an equivalent statement, which connects to 2s-sparse observability in the centralized framework in certain scenarios, such that the sufficient condition is feasible.

The remainder of the paper is organized as follows: Section \ref{sec_formulation} is on preliminaries  and problem formulation. Section \ref{sec_filter} considers the secure distributed filter.  Section \ref{sec_analysis} provides the performance analysis for the filter. Section \ref{sec_simu} gives the numerical simulation results. The conclusions of this paper are given in Section \ref{sec_conclusion}.

\section{Problem Formulation}\label{sec_formulation}
\subsection{Notations}
The superscript ``T" represents the transpose. $\mathbb{R}^{n\times m}$ is the set of real matrices with $n$ rows and $m$ columns.  $\mathbb{R}^n$ is the $n$-dimensional Euclidean space.
$I_{n}$ stands for the $n$-dimensional square identity matrix. 
$\textbf{1}_N$ stands for the $N$-dimensional vector with all elements being one. 
$\diag\{\cdot\}$   represents the diagonalization operator.  
$A\otimes B$ is the Kronecker product of $A$ and $B$.  $\norm{x}_2$ is the 2-norm of a vector $x$. $\norm{A}_2$ is the induced 2-norm, i.e., $\norm{A}_2=\sup\limits_{x\neq 0}\frac{\norm{Ax}_2}{\norm{x}_2}$.  
$\lambda_2(A)$ and $\lambda_{max}(A)$ are the second minimal eigenvalue and maximal eigenvalue of $A$, respectively.
\subsection{Graph Preliminaries}
In an undirected graph $\mathcal{G=(V,E)}$, $\mathcal{V}$ stands for the set of  nodes, $\mathcal{E}\subseteq \mathcal{V}\times \mathcal{V}$ is the  set of edges.  If there is an edge $(i,j)\in \mathcal{E}$, node $i$ can exchange information with node $j$, and node $j$  is called a  neighbor of node $i$. 
Let the neighbor set of agent $i$ be $\mathcal{N}_{i}:=\{j\in\mathcal{V}|(i,j)\in \mathcal{E}\}$.
The graph $\mathcal{G}$ is  connected if for any pair of nodes $(i_{1},i_{l})$, there exists a  path from $i_{1}$ to $i_{l}$ consisting of edges $(i_{1},i_{2}),(i_{2},i_{3}),\ldots,(i_{l-1},i_{l})$.  
$\mathcal{L}$ is the Laplacian matrix whose definition is referred to \cite{Mesbahi2010Graph}.  
On the connectivity of a graph, we have
\begin{proposition}\cite{Mesbahi2010Graph}\label{thm_graph}
	The undirected graph $\mathcal{G}$ is	 connected if and only if $\lambda_{2}(\mathcal{L})>0$.
\end{proposition}

\subsection{System model}
Consider the following plant observed by $N$ agents (e.g., sensors),
\begin{equation}\label{eq_system}
\begin{split}
x(t+1)&=Ax(t)+w(t)\\
y_i(t)&=C_ix(t)+v_i(t)+a_i(t),i=1,\dots,N,
\end{split}
\end{equation}
where $x(t)\in\mathbb{R}^n$ is the unknown system state, $w(t)\in\mathbb{R}^n$ is the process  noise,  $v_i(t)\in\mathbb{R}$ is the observation  noise, and $a_i(t)\in\mathbb{R}$ is the  attack signal inserted by some malicious attacker, all at time $t$. $y_i(t)\in\mathbb{R}$ is the observation of agent $i$. Moreover, $A\in\mathbb{R}^{n\times n}$ is the system state transition matrix, and $C_i\in\mathbb{R}^{1\times n}$ is the observation vector of agent $i$.
\begin{remark}
	The essential problem is to study the influence of  scalar attack signal to the estimation performance with certain number of compromised observation elements, like \cite{chen2018topology}. Thus, we consider the observation equation with scalar outputs for each agent. This conforms with the centralized framework, where each row vector of centralized observation matrix stands for the observation vector of one agent.
\end{remark}

\begin{definition}(One-step Collective Observability)\label{def_one_obser}
	The system \eqref{eq_system} is called one-step collectively observable if $\sum_{i=1}^NC_i^TC_i$ is a positive definite matrix. 
\end{definition}
\begin{remark}
	On the relation between one-step collective observability, which requires $N\geq n$, and $n$-step collective observability (i.e., $(A,C)$ is observable, where $C=[C_1^T,\dots,C_N^T]$): If $A$ is a diagonal matrix such as $A=I_n$, the two definitions are equivalent. For general system matrices, $n$-step  collective observability is milder than the one-step collective observability. Notice also that one-step collective observability does not mean local observability, i.e., $(A,C_i)$ could be unobservable or undetectable, $\forall i=1,\dots,N$. 
\end{remark}

In this paper, the following assumptions are in need.
\begin{assumption}\label{ass_bounds}
	The following conditions hold
	\begin{align*}
	&\norm{A}_2=a\geq 1,\norm{w(t)}_2\leq b_{w},\norm{v_i(t)}_2\leq b_{v},\\
	&\norm{\hat x_i(0)-x(0)}_2\leq \eta_{i}\leq\eta_0, i=1,\cdots,N,
	\end{align*}
	where $\hat x_i(0)$ is the estimate of $x(0)$ by agent $i$. Besides, the bounds are known to each agent.
\end{assumption}

\begin{assumption}\label{ass_C}
	The system \eqref{eq_system} is one-step collectively observable, i.e., $\sum_{i=1}^NC_i^TC_i\succ0$. 
	The observation vector $C_i$ is  normalized, i.e., $\norm{C_{i}}_2=1, i=1,\cdots,N$. 
\end{assumption}

\begin{assumption}\label{ass_graph}
	The communication graph $\mathcal{G}=(\mathcal{V},\mathcal{E})$ is undirected and connected, where $\mathcal{V}=\{1,2,\dots,N\}$
\end{assumption}
\begin{remark}
	To design a non-trivial filtering algorithm with guaranteed bounded estimator error, we assume $\norm{A}_2\geq1$ in  Assumption \ref{ass_bounds}.    Otherwise, one can easily design a filter such that  estimation errors keep bounded. The proposed methods and results also apply to the case where $\norm{A}_2<1.$  Assumption \ref{ass_C} requires a collective observability condition utilized in the existing literature on distributed estimation \cite{Khan2014Collaborative,kar2011convergence,kar2013distributed}. The normalized observation vectors 	
	can be obtained by reconstructing the system \eqref{eq_system}.  Different from \cite{mitra2016secure} requiring some robustness of communication graph, the connectivity of  Assumption \ref{ass_graph} is a standard condition for distributed estimation. If  the graph  is not connected, the problem can be studied for the connected subgraphs separately. 
\end{remark}

A typical distributed filtering problem is to design an online filter or state estimator for each agent (e.g., agent $i$) to estimate the system state $x(t)$ by employing the known local noisy observations $\{y_i(l)\}_{l=1}^{t}$ and the messages received from neighboring agents.
However, if  observations of some agents are compromised by a malicious attacker,  the observation quality may be tremendously affected, which will  bring big challenges in design and analysis of distributed filtering algorithms. In the following, we introduce the  attack model.
\subsection{Attack model}
To deteriorate the estimation performance of filtering algorithms, the malicious attacker aims to persistently destroy the observation data of some targeted agents. However, due to  resource limitation, the attacker has limited power to  attack the set of agents. Assume that 
the set of compromised agents  is fixed over time, and consists of no more than $s$ agents.
Since the knowledge of the attacker makes a big difference to its ability in deteriorating the estimation performance, we assume the following knowledge scope of the attacker. 
\begin{assumption}\label{ass_attacker}
	The attacker has  full knowledge on the system (\ref{eq_system}), the network topology,  and the filter of all agents. Furthermore, 
	the observation $y_i(t)$ can be arbitrary for a compromised agent $i$.
\end{assumption}
Under Assumption \ref{ass_attacker}, we have
\begin{equation}\label{attack_input}
\begin{split}
&a_i(t)\in R, i\in\mathcal{A}, \text{ with }|\mathcal{A}|\leq s\\
&a_i(t)=0, i\in \mathcal{N}=\mathcal{V}-\mathcal{A},\forall t\in\mathbb{N},
\end{split}
\end{equation}
where $\mathcal{A}$ is the  set  of agents whose observations are compromised by the malicious attacker. $\mathcal{N}$ is the set of normal agents without being affected by the attacker.  Note that the sets $\mathcal{A}$ and $\mathcal{N}$ are unknown to each agent.

\begin{remark}
	In Assumption \ref{ass_attacker}, we consider the worst scenario on compromised observations that the attacker can access the full information without requiring any concrete attack models, which is more general than  results in the existing literature \cite{deghat2019detection}. 
\end{remark}

We further have the following definitions.
\begin{definition}($s$-sparse observability)\label{def_sparse}
	The linear system defined by (\ref{eq_system})  is said to be $s$-sparse observable if for every set $\Gamma\subseteq \{1,\dots,N\}$ with $|\Gamma|=s$, the pair $(A,C_{\bar \Gamma})$ is observable, where $C_{\bar \Gamma}$ is the remaining matrix by removing $C_{j},j\in\Gamma$ from $[C_1^T,C_2^T,\dots,C_N^T]$.
\end{definition}

\begin{definition}(One-step $s$-sparse observability)\label{def_sparse_one}
	The linear system defined by (\ref{eq_system}) is said to be one-step $s$-sparse observable if for every set $\Gamma\subseteq \{1,\dots,N\}$ with $|\Gamma|=s$, the pair $C_{\bar \Gamma}^TC_{\bar \Gamma}=\sum_{i=1,i\notin\Gamma}^{N}C_{i}^TC_{i}\succ0$, where $C_{\bar \Gamma}$ is the remaining matrix by removing $C_{j},j\in\Gamma$ from $[C_1^T,C_2^T,\dots,C_N^T]$.
\end{definition}

\begin{remark}
	
	Definition \ref{def_sparse} and Definition \ref{def_sparse_one} correspond to the $n$-step (collective) observability (i.e., $(A,C)$ is observable) and one-step (collective) observability in Definition \ref{def_one_obser}.  If the system matrix $A$ is diagonal,	Definition \ref{def_sparse} and Definition \ref{def_sparse_one} are equivalent.  	 In the centralized framework, if the observations of $s$ agents are compromised, the system should be $2s$-sparse observable to guarantee the effective estimation of system state \cite{shoukry2016event}.
\end{remark}
%


\subsection{Problems of Interest}
We mainly consider the following problems in this paper.

1) How to design secure distributed filter  for each agent by employing the local noisy observations potentially compromised by the malicious attacker?


2) What   conditions can guarantee the bounded estimation error of  the distributed filter in presence of the attacker (\ref{attack_input}). How can we quantify the estimation performance of the distributed filter?

\section{Secure Distributed Filter:SDCF}\label{sec_filter}
In this section, we will design  a secure distributed filter for each agent.

We consider the filtering algorithm with two stages, namely, observation update and consensus. In the stage of local observation update,  we design a saturation-like scheme to utilize the observation $y_{i}(t)$ as follows
\begin{align}\label{alg_update}
\tilde x_{i}(t)=&A\hat x_{i}(t-1)+ k_{i}(t)C_i^T(y_{i}(t)-C_iA\hat x_{i}(t-1)), 
\end{align}
where 
\begin{align}\label{eq_K}
k_{i}(t)=\begin{cases}
1,\text{ if } |y_{i}(t)-C_{i}A\hat x_{i}(t-1)|\leq\beta,\\
\frac{\beta}{|y_{i}(t)-C_{i}A\hat x_{i}(t-1)|}, \text{ otherwise}.
\end{cases}
\end{align}

Different from the gain designs of common filters or state estimators, the gain $k_{i}(t)$ in this work is related to the value of innovation (i.e., $y_{i}(t)-C_iA\hat x_{i}(t-1)$). 
The design of $k_{i}(t)$ in (\ref{eq_K}) makes sense, since if the estimation innovation  is very large, the observation $y_i(t)$ is more likely to be compromised. 
By the designed gain $k_{i}(t)$, we have $|k_{i}(t)(y_{i}(t)-C_iA\hat x_{i}(t-1))|\leq\beta$, which guarantee that the attacker has limited influence to the local update stage of the filter. 

In the consensus stage, we suppose that each agent can communicate with its neighbors for $L\geq 1$ times between two time instants.
%
For $l=1,2,\dots,L,$
\begin{align}\label{alg_consensus}
\hat x_{i,l}(t)=\hat x_{i,l-1}(t)-\alpha\sum_{j\in\mathcal{N}_{i}}(\hat x_{i,l-1}(t)-\hat x_{j,l-1}(t)),
\end{align}
with $\hat x_{i,0}(t)=\tilde x_{i}(t)$ and we denote  $\hat x_{i}(t)=\hat x_{i,L}(t).$ For each communication, agent $j$ will transmit its estimate $\hat x_{j,l-1}(t)$ to its neighbors, $l=1,\dots,N.$
\begin{remark}
	The parameter $\beta$ in \eqref{eq_K} reflects the usage tradeoff between normal observations and compromised observations.	
	If $\beta$ is very large, then almost all normal observations will be utilized without scaling. But, it will give much   space that the attacker can use to deteriorate the estimation performance. If $\beta$ is very small, although the most possible attack signals may be filtered by the designed gain $k_{i}(t)$, many normal observations will contribute little to the estimation performance. As a result, the filtering error of each agent will probably be divergent.  
	The condition on  $\beta$  will be discussed in next section. 
\end{remark}
\begin{remark}
	The term $\alpha\sum_{j\in\mathcal{N}_{i}}(\hat x_{i,l-1}(t)-\hat x_{j,l-1}(t))$ is to make the agents reach consensus.
	The consensus step is vital to guarantee bounded estimation error  of distributed filters especially for the case that each subsystem is not observable (i.e., $(A,C_i)$ is not observable). 
	The parameter $\alpha$ can increase the consensus speed if it is well designed.	
	It can be proven that if the consensus step $L$ goes to infinity and the parameter $\alpha$ is properly designed, then the estimates $\{\hat x_{i}(t)\}_{i=1}^N$ will converge to the same vector.
	However, the consensus step $L$ is not required to approximate  infinity in this work. The requirement of the step $L$ and the design of  $\alpha$, related with the system structure and performance demand, is  given in next section. 
\end{remark}

By (\ref{alg_update}), (\ref{eq_K}) and (\ref{alg_consensus}), we obtain the secure distributed consensus filter (SDCF) in Algorithm \ref{alg:A}.
\begin{algorithm}
	\caption{Secure Distributed Consensus Filter (SDCF):}
	\label{alg:A}
	\begin{algorithmic}[1]
		\STATE {\textbf{Update:} Agent $i$ uses its own observation to update the estimate}\\
		$\tilde x_{i}(t)=A\hat x_{i}(t-1)
		+ k_{i}(t)C_i^T(y_{i}(t)-C_iA\hat x_{i}(t-1))\nonumber$\\        
		\begin{flushleft}
			$k_{i}(t)=\min\{1,\frac{\beta}{|y_{i}(t)-C_{i}A\hat x_{i}(t-1)|}\}$, 
		\end{flushleft}
		\STATE {\textbf{Consensus for $L$ steps: $\hat x_{i,0}(t)=\tilde x_{i}(t)$}}\\
		For $l$th consensus, $l=1,\dots,L$:\\
		\quad For $i$th agent, $i=1,\dots,N$:\\
		\qquad Agent $i$ receives $\hat x_{j,l-1}(t)$, $j=1,\dots,\mathcal{N}_{i}$, then \\
		$\qquad\hat x_{i,l}(t)=\hat x_{i,l-1}(t)-\alpha\sum_{j\in\mathcal{N}_{i}}(\hat x_{i,l-1}(t)-\hat x_{j,l-1}(t))$\\	
		\quad end\\
		end\\
		\STATE {\textbf{Output step: $\hat x_{i}(t)=\hat x_{i,L}(t).$} }
	\end{algorithmic}
\end{algorithm}

\section{Performance analysis}\label{sec_analysis}
In this section, we will focus on performance analysis of the proposed SDCF algorithm.
Specifically, we will study the conditions to guarantee the boundedness of estimation error, and quantify the estimation performance under compromised observations.

For convenience, we denote
\begin{align}\label{eq_denotations}
X(t)&=\textbf{1}_N\otimes  x(t),\nonumber\\
Y(t)&=\left[y_1^T(t),\dots,y_N^T(t)\right]^T,\nonumber\\
V(t)&=\left[v_1^T(t),\dots,v_N^T(t)\right]^T,\nonumber\\
\hat X(t)&=\left[\hat x_1^T(t),\dots,\hat  x_N^T(t)\right]^T,\\
\bar C&=\diag\{C_1,\dots,C_N\},\nonumber\\
\bar K(t)&=\diag\{k_1(t),\dots,k_N(t)\},\nonumber\\
P_{Nn}&=\frac{1}{N}(\textbf{1}_N\otimes  I_n)(\textbf{1}_N\otimes  I_n)^T\nonumber.
\end{align}
By (\ref{alg_update}), (\ref{alg_consensus}) and (\ref{eq_denotations}), for $i=1,\dots,N$, we obtain the compact form of recursive state estimates of SDCF in the following 
\begin{align}\label{eq_recursive}
\hat X(t)=&\left(I_{Nn}-\alpha(\mathcal{L}\otimes I_n)\right)^L\bigg[(I_N\otimes A)\hat X(t-1)\nonumber\\
&+ \bar C^T \bar K(t)(Y(t)-\bar C(I_N\otimes A)\hat X(t-1))\bigg],
\end{align}
where $\mathcal{L}$ is the Laplacian matrix.
Define $E(t)=\hat X(t)-\textbf{1}_N\otimes x(t)$, noting $(\mathcal{L}\otimes I_n)(\textbf{1}_N\otimes x(t))=0$, then we obtain the error dynamics as follows
\begin{align}\label{eq_recursive_error}
&E(t)\nonumber\\
=&\left(I_{Nn}-\alpha(\mathcal{L}\otimes I_n)\right)^L\bigg[(I_{Nn}-\bar C^T\bar K(t)\bar C)(I_N\otimes A)E(t-1)\nonumber\\
&+(\bar C^T\bar K(t)\bar C-I_{Nn})(I_{N}\otimes w(t-1))+ \bar C^T\bar K(t)V(t)\nonumber\\
&+ \bar C^T\bar K(t)a(t)\bigg],
\end{align}
where $a(t)=\diag\{a_1(t),a_2(t),\dots,a_N(t)\}.$
\begin{remark}
	Since the filtering gain $\bar K(t)$ is related to the state estimates and  potential compromised observations, the common stability analysis approaches, such as Lyapunov methods, may not be directly utilized to analyze the stability or boundedness of estimation error $E(t)$ by its dynamics (\ref{eq_recursive_error}). 
	This is the main challenge for the problem of distributed recursive filter under compromised observations.
\end{remark}
\subsection{Boundedness of estimation error}
In this subsection, we will study the conditions to guarantee the boundedness of estimation error for the SDCF  in Algorithm \ref{alg:A}.  
Denote $\lambda_0:=\lambda_{min}\left( \sum_{i\in\mathcal{N}^*}C_{i}^TC_{i}\right)$, where $\mathcal{N}^*$ is the agent set such that  $\lambda_{min}\left( \sum_{i\in\mathcal{N}}C_{i}^TC_{i}\right)$ is minimal within all sets $\{\mathcal{N}\}$ obtained by removing any $|\mathcal{A}|$ agents from $\mathcal{V}$. Besides, for convenience, we give the following notations
\begin{align}\label{denotations_thm}
&\gamma=\frac{\lambda_{max}(\mathcal{L})-\lambda_2(\mathcal{L})}{\lambda_{max}(\mathcal{L})+\lambda_2(\mathcal{L})},\nonumber\\
& p^*_0=a\gamma^L\sqrt{N} \eta_0+\frac{\sqrt{N}\beta\gamma^L}{1-a\gamma^L},\nonumber\\
&k^*=\min\{1,\frac{\beta}{a(p_0^*+\eta_0)+b_{w}+b_{v}}\},\nonumber\\
&\mu_0=a\left(1-\frac{k^*}{N}\lambda_0\right),\\
&Q_0=(1-\frac{|\mathcal{A}|}{N})(b_{w}+b_{v}+ap_0^*)+b_{w},\nonumber\\
&\vartheta_0=1-\frac{Q_0}{\eta_0}\left(1-\frac{\beta|\mathcal{A}|}{N\eta_0}\right)^{-1},\nonumber\\
&m_0=\vartheta_0\left(1-\frac{\beta|\mathcal{A}|}{N\eta_0}\right)\left(1-\frac{k^*\lambda_0}{N}\right)^{-1}\nonumber.
\end{align}	
On the boundedness of estimation error by SDCF in Algorithm \ref{alg:A}, we have the following result.
\begin{theorem}\label{thm_stability}
	Let Assumptions \ref{ass_bounds} - \ref{ass_graph} hold and $ \alpha=\frac{2}{\lambda_2(\mathcal{L})+\lambda_{max}(\mathcal{L})}$.
	If there exist a set of scalars $L>0$, $\beta>0$, $\eta_0> 0$, such that
	\begin{align}\label{condition_thm}
	1\leq a< \min\left\{m_0,\gamma^{-L}\right\},
	\end{align}
	then  the estimation error of $\text{SDCF}(L,\beta)$, i.e., $e_{i}(t)=\hat x_{i}(t)-x(t)$,  $\forall i\in\mathcal{V}$, satisfies 
	\begin{align}
	\lim\limits_{t\rightarrow \infty}\norm{e_{i}(t)}_2\leq \frac{NQ_0+|\mathcal{A}|\beta}{N(1-\mu_0)}+\frac{\sqrt{N}\beta\gamma^L}{1-a\gamma^L}<\infty.
	\end{align}
\end{theorem}
\begin{remark}
	The parameters $\beta$, $L$ are given in the implementation of algorithm. Although $\eta_0$ is a bound of initial estimation error, we can adjust it  bigger to meet the requirement.
\end{remark}
\begin{remark}
	Theorem \ref{thm_stability} shows that by taking proper parameters $L,\beta,\eta_0$, the SDCF can guarantee the boundedness of estimator error for a class of unstable dynamics. The condition \eqref{condition_thm} can be examined offline with global knowledge to provide the parameter design.
\end{remark}


\subsection{Feasibility of Condition \eqref{condition_thm}}
Since the condition \eqref{condition_thm} is complex, its feasiblity needs to be testified, i.e., whether there exists a set of positive parameters $L,\beta,\eta_0$ such that   \eqref{condition_thm} is satisfied. In this subsection, we study the feasibility of \eqref{condition_thm}. 
\begin{theorem}\label{thm_iff}
	Condition (\ref{condition_thm}) has a feasible solution on $\beta,\eta_0$ and $L$,
	if and only if
	\begin{align}\label{eq_iff}
	\lambda_0>|\mathcal{A}|.
	\end{align}
\end{theorem}
\begin{remark}
	Recall $\lambda_0:=\lambda_{min}\left( \sum_{i\in\mathcal{N}^*}C_{i}^TC_{i}\right)$, which reflects the one-step sparse observability of the system by removing any $|\mathcal{A}|$ agents.
	Since the compromised subset of agents is fixed over time,  
	we can calculate $\lambda_0$ and compare it with $|\mathcal{A}|$.
\end{remark}

The direct relationship between \eqref{eq_iff} and the one-step sparse observability is given in the following.
\begin{lemma}\label{lem_iff}
	A necessary condition to guarantee $\lambda_0>s:=|\mathcal{A}|$ is that the system (\ref{eq_system}) is one-step $2s$-sparse observable. If the observation vectors are orthogonal and $A$ is a diagonal matrix, then one-step $2s$-sparse observability is also a sufficient condition to guarantee $\lambda_0>|\mathcal{A}|$.
\end{lemma}

\begin{remark}
	From Lemma \ref{lem_iff} and Theorem \ref{thm_iff}, on Algorithm \ref{alg:A}, we have that if the observations of any $s$ agents are under attacks, the system (\ref{eq_system}) should be one-step $2s$-sparse observable  to achieve the effective estimation of system state. For the case that  $A$ is a diagonal matrix,  the condition \eqref{eq_iff} conforms to the centralized framework that the system is $2s$-sparse observable \cite{shoukry2016event}.
\end{remark}

\subsection{Proof of Theorem \ref{thm_stability}}\label{subs_proof}
In this subsection, we provide  the proof of Theorem \ref{thm_stability}.
\begin{lemma}\label{lem_mut}
	The following equation holds
		\begin{align*}
		&\left(I_{Nn}-\alpha(\mathcal{L}\otimes I_n)-P_{Nn}\right)^LM_{t}\nonumber\\
		=&\left(I_{Nn}-\alpha(\mathcal{L}\otimes I_n)\right)^LM_{t},
		\end{align*}
	where $M_{t}=(I_N\otimes A)(\hat X(t)-\textbf{1}_N\otimes x_{avg}(t))$, $\hat x_{avg}(t)=\frac{1}{N}\sum_{i=1}^N\hat x_{i}(t)$.
\end{lemma}

The following lemma provides an optimal design for the consensus parameter $\alpha$, which is related to the graph topology. 
\begin{lemma}\label{lem_alpha}
	Under Assumption \ref{ass_graph}, the following result holds
	\begin{align}
	\min_{\alpha}\norm{I_{Nn}-\alpha(\mathcal{L}\otimes I_n)-P_{Nn}}_2
	=&\frac{\lambda_{max}(\mathcal{L})-\lambda_2(\mathcal{L})}{\lambda_{max}(\mathcal{L})+\lambda_2(\mathcal{L})}<1,\nonumber
	\end{align}
	with the optimal  solution $\alpha^*=\frac{2}{\lambda_2(\mathcal{L})+\lambda_{max}(\mathcal{L})}.$
\end{lemma}

 The following lemma studies the influence of consensus step $L$ to the observation innovation part of  (\ref{eq_recursive}). 
\begin{lemma}\label{lem_consensus}
	Let $\alpha=\frac{2}{\lambda_2(\mathcal{L})+\lambda_{max}(\mathcal{L})}$, then the following result holds
	\begin{align*}
	&\norm{(I_{Nn}-P_{Nn}) \left(I_{Nn}-\alpha(\mathcal{L}\otimes I_n)\right)^L\bar Y(t)}_2\leq \sqrt{N}\gamma^L\beta,
	\end{align*}
	where $\bar Y(t)=\bar C^T\bar  K(t)(Y(t)-\bar C(I_N\otimes A)\hat X(t-1)).$
\end{lemma}

Let $e_i(t)$ be the estimation error of agent $i$ by the filter in Algorithm \ref{alg:A}, i.e., $e_i(t)=\hat x_{i}(t)-x(t)$. 
Then we have $e_i(t)=\hat x_{i}(t)-x(t)=\tilde e(t)+\bar e_i(t)$, 
where $\bar e_i(t):=\hat x_{i}(t)-\hat x_{avg}$, and $\tilde e(t)=\hat x_{avg}-x(t),$ and $\hat x_{avg}(t):=\frac{1}{N}\sum_{i=1}^N\hat x_{i}(t)$. 
The idea to analyze the boundedness of estimation error $e_i(t)$ is to find conditions that can guarantee the boundedness of $\bar e_i(t)$ and $\tilde e(t)$ simultaneously. As a result, the boundedness of $e_i(t)$ can be guaranteed. In the following, Lemma \ref{prop_consensus} and Lemma \ref{prop_avg} study the   boundedness of $\bar e_i(t)$ and $\tilde e(t)$, respectively.
\begin{lemma}\label{prop_consensus}
	Consider Algorithm \ref{alg:A} with $L\geq 1$, and let Assumptions \ref{ass_bounds} - \ref{ass_graph} hold.
	If $ \alpha=\frac{2}{\lambda_2(\mathcal{L})+\lambda_{max}(\mathcal{L})}$, and $\norm{A}_2=a<\gamma^{-L},$
	then 
	\begin{align}\label{eq_error_norm}
	\norm{\bar e_i(t)}_2
	\leq p^*(L,t),
	\end{align}
	where $p^*(L,t)=(a\gamma^L)^t\sqrt{N}\eta_0+\sqrt{N}\beta\gamma^L\frac{1-a^{t-1}\gamma^{L(t-1)}}{1-a\gamma^L}$.
	Furthermore, $\sup_{t\geq 1}\{p^*(L,t)\}\leq a\gamma^L\sqrt{N} \eta_0+\frac{\sqrt{N}\beta\gamma^L}{1-a\gamma^L}\triangleq p^*_0<\infty,$ and
	\begin{equation}\label{eq_asymp}
	\begin{split}
\lim\limits_{L\rightarrow \infty}p^*(L,t)&=0,\\
\lim\limits_{t\rightarrow\infty} p^*(L,t)&=\frac{\sqrt{N}\beta\gamma^L}{1-a\gamma^L}<\infty.
	\end{split}
	\end{equation}
\end{lemma}
\begin{remark}
	Lemma \ref{prop_consensus} shows that the error between each state estimate and the average estimates can be upper bounded by $p^*(L,t)$, which is uniformly upper bounded by a constant scalar $p^*_0$ and has some asymptotic properties w.r.t. consensus step $L$ and time $t$. 
\end{remark}



\begin{lemma} \label{prop_avg}	
	Consider Algorithm \ref{alg:A} with $L\geq 1$, and assume that Assumptions \ref{ass_bounds} - \ref{ass_graph} hold,  
	$ \alpha=\frac{2}{\lambda_2(\mathcal{L})+\lambda_{max}(\mathcal{L})}$, and $\norm{A}_2=a<\gamma^{-L}$.
	If
	\begin{align}\label{eq_all_condition}
	\frac{|\mathcal{A}|\beta+NQ_0}{N\eta_0}\leq 1-\mu_0,
	\end{align}
	then 
	\begin{align}\label{eq_asym_avg}
	\lim\limits_{t\rightarrow \infty}\norm{\tilde e(t)}_2\leq\frac{NQ_0+|\mathcal{A}|\beta}{N(1-\mu_0)}.
	\end{align}
	%
	%
\end{lemma}
\begin{remark}
	Lemma \ref{prop_avg} provides the sufficient condition to guarantee the boundedness of network tracking error (i.e., $\hat x_{avg}(t)-x(t)$). For given $b_{w}, b_{v}, \lambda_0$ and $a$, we can design $\beta$ and $L$ based on the condition \eqref{eq_all_condition} to guarantee \eqref{eq_asym_avg}.

\end{remark}

From \eqref{condition_thm} and $a\gamma^L<1$, we can have the condition \eqref{eq_all_condition}. By Lemma \ref{prop_consensus}, Lemma \ref{prop_avg} and the notations in \eqref{denotations_thm}, the conclusion of Theorem \ref{thm_stability} holds.

\section{Simulation Results}\label{sec_simu}
In this section, we  carry out a numerical simulation to show the effectiveness of the proposed algorithm.

\begin{figure}[htp]
	\centering
	\includegraphics[scale=0.5]{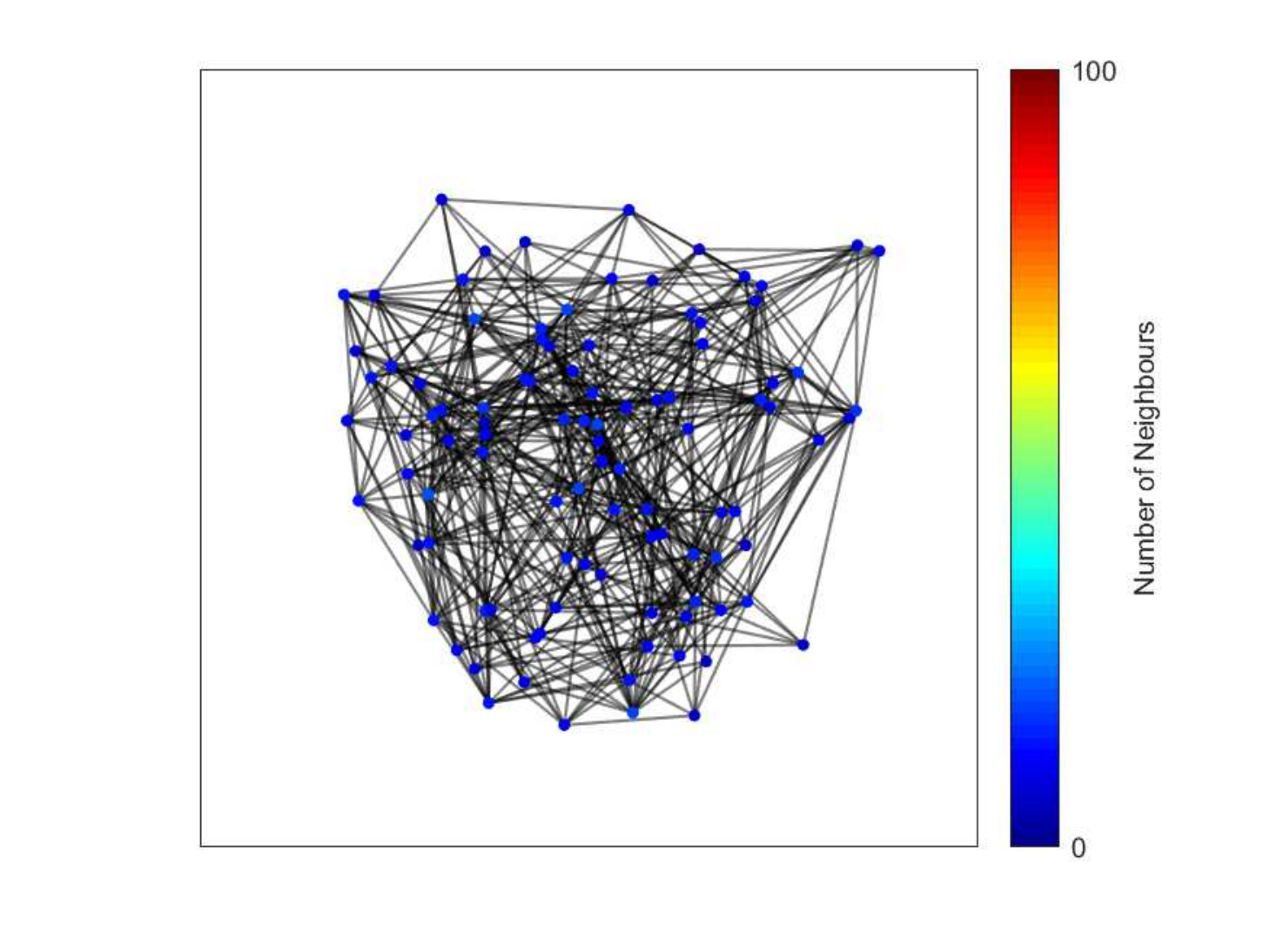}
	\caption {A random sparse connected graph with 100 nodes.}
	\label{fig:networks}
\end{figure}

Regarding the system \eqref{eq_system}, we assume $A=\left[\begin{smallmatrix}
1.01& 0.1\\
0.1&1.1
\end{smallmatrix}
\right]$ with $\norm{A}_2=1.16$. The observation vectors are randomly selected from the  set $\left\{C_1=[1,0],C_2=[0,1],C_3=[\frac{\sqrt{2}}{2},\frac{\sqrt{2}}{2}]\right\}$. The process noise $w(t)$ and observation noises $v_i(t),i=1,\dots,N$,  all follow the uniform distribution between $[0,1]$. 
The bounds are assumed to be $b_v=1, b_w=1,\eta_i=1,i=1,\dots,N.$
We suppose the time $t=[0,100]$ with sampling interval $1$. 
The sparse network given in Fig. \ref{fig:networks} has $N=100$ nodes with  $\lambda_2(\mathcal{L})=4.1$ and $\lambda_{max}(\mathcal{L})=21.3$. We choose $\beta=3$, the times of Monto Carlo experiments is $100$. Suppose that the attacker will insert the signal $a_i(t)=2(C_ix(t)+v_i(t))$  if   agent $i$ is compromised. 

\begin{figure}[htp]
	\centering
	\includegraphics[scale=0.5]{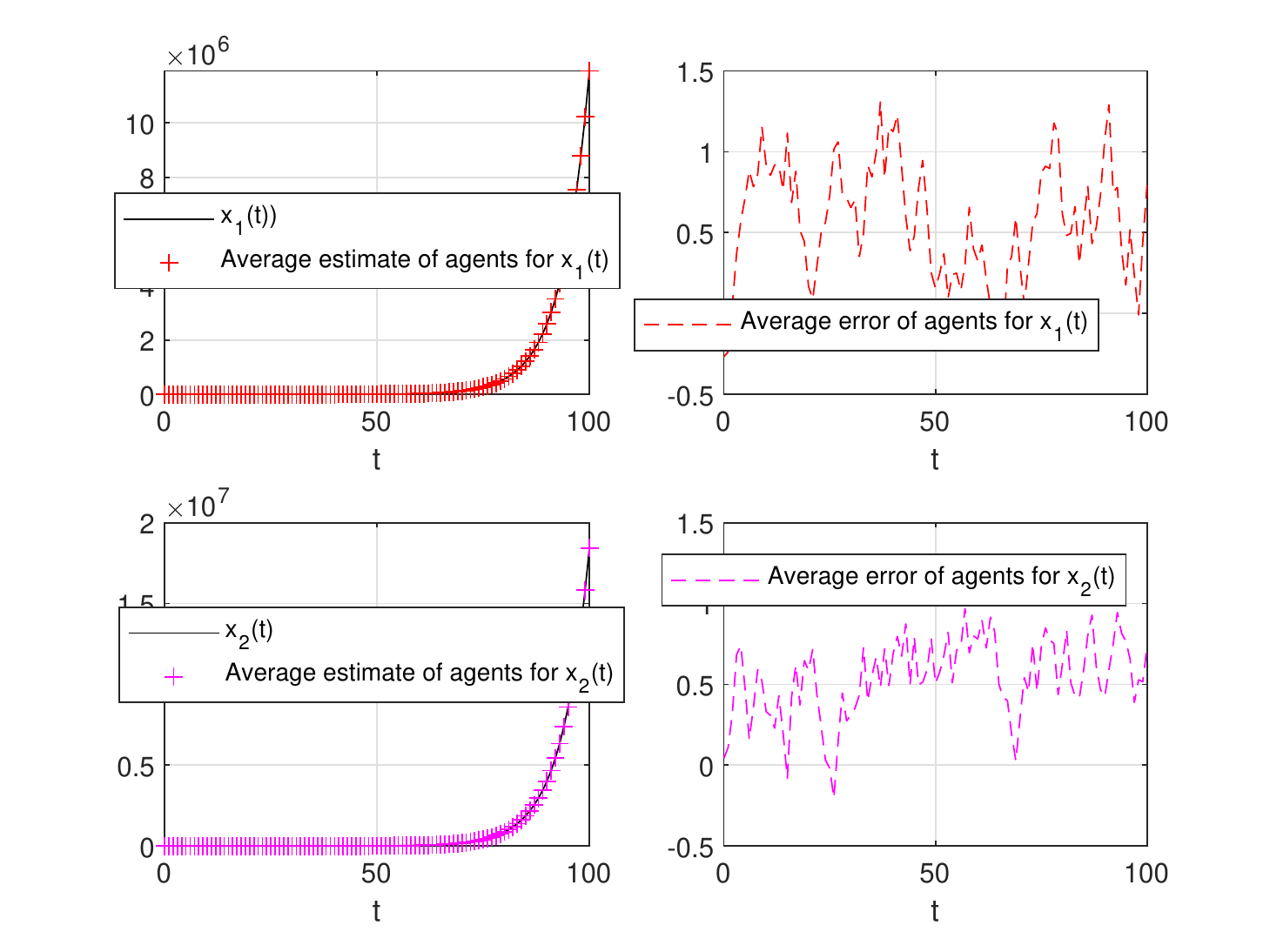}
	\caption {Network tracking performance for each element over one realization.}
	\label{fig:tracking}
\end{figure}

\begin{figure}[htp]
	\centering
	\includegraphics[scale=0.5]{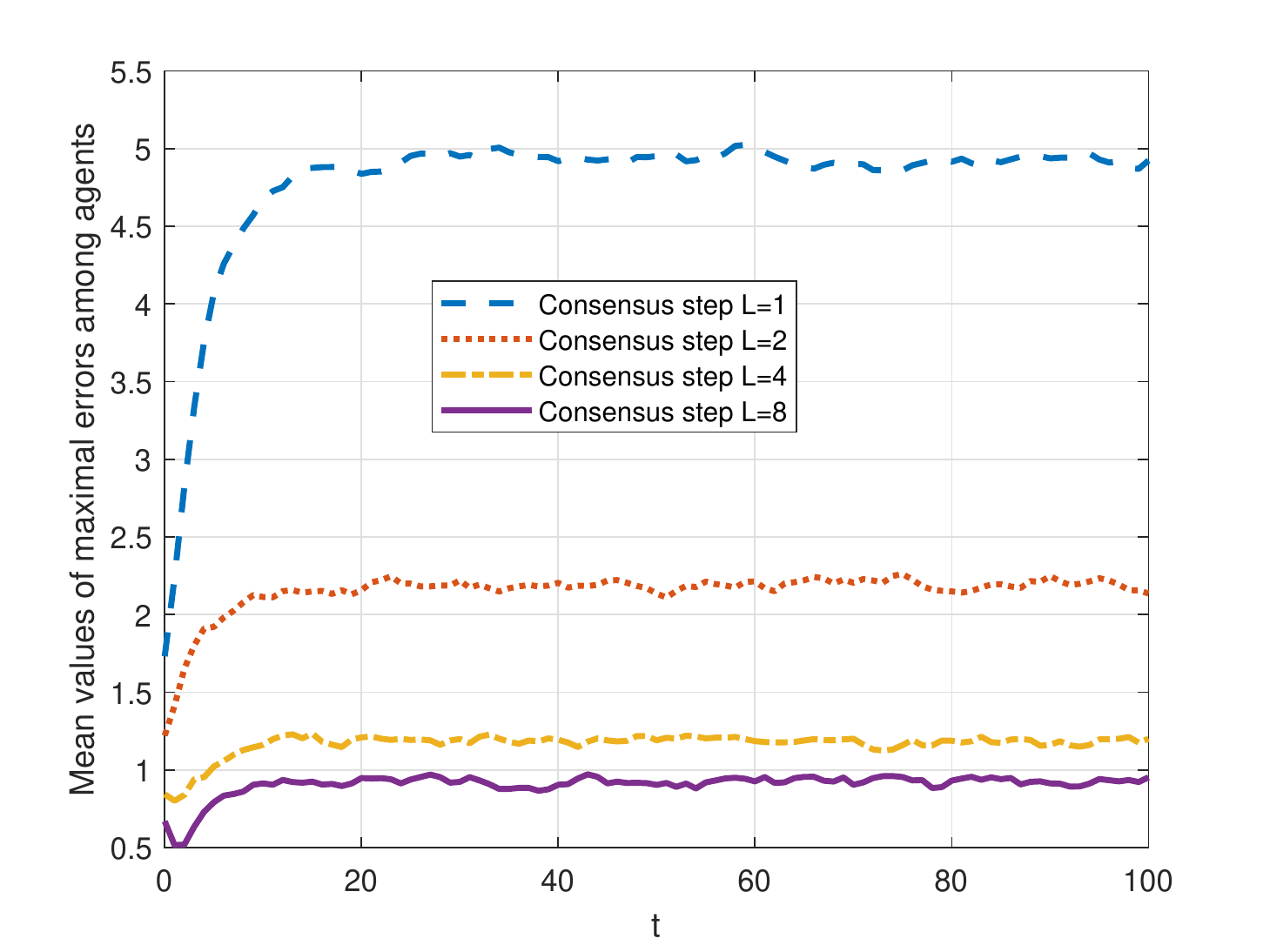}
	\caption {The influence of consensus step to  error norm dynamics.}
	\label{fig:consensus}
\end{figure}

\begin{figure}[htp]
	\centering
	\includegraphics[scale=0.5]{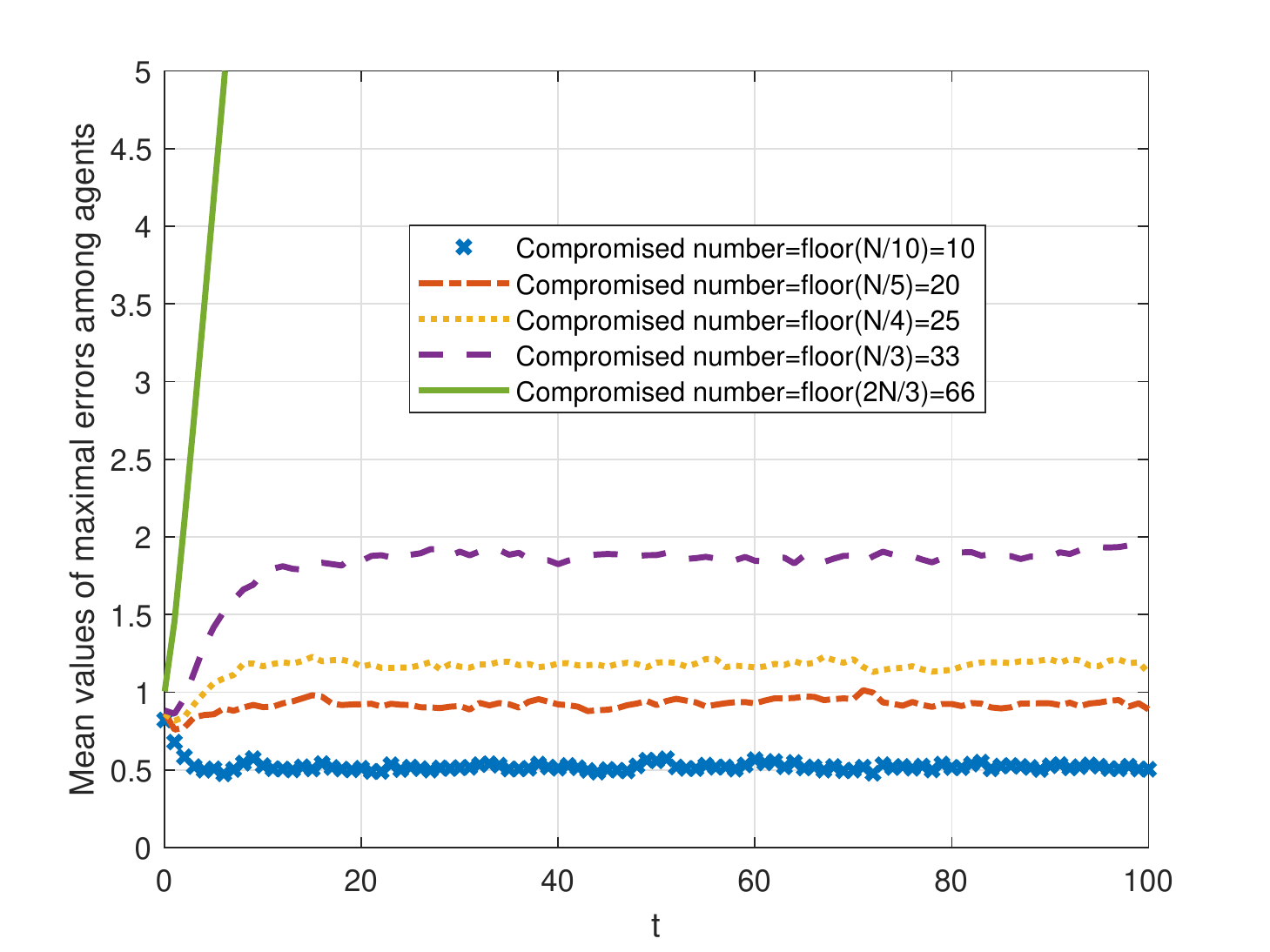}
	\caption {The influence of compromised agent number to error norm dynamics.}
	\label{fig:attack}
\end{figure}

We carry out the numerical simulation by employing the SDCF to study its estimation performance under the above setting. For one realization with consensus step $L=8$ and the number of compromised agents $25$, we obtain the network tracking performance in Fig. \ref{fig:tracking}. It   shows  that each element of the system state, i.e., $x_1(t)$ and $x_2(t)$, can be well estimated by agents over the network with small bounded estimation error. The influence of consensus step $L$ to the mean values (averaged by 100) of maximal errors among all agents is studied in Fig. \ref{fig:consensus} with the number of compromised agents $25$, which shows that a bigger consensus step can lead to smaller estimation error. In Fig. \ref{fig:attack} with $L=4$, we investigate the influence of compromised agent number to the estimation error. We see that with the increasing of compromised number, the estimation errors will become larger, and even diverge when the number is 66. The phenomena  conform with former  analysis, since not enough information can support an effective estimator if too many agents are compromised.   Based on the above results, the utility of the proposed SDCF is validated.  
\section{Conclusions}\label{sec_conclusion}
This paper studied the secure distributed filtering problem for linear time-invariant systems with bounded noises and unstable dynamics under compromised observations. We considered a general case that a malicious attacker can compromise a subset of  agents and manipulate the observations arbitrarily.
First, we proposed a consensus-based distributed filter by employing a  saturation-like scheme, which gives a small gain if the innovation is too large.
Then, we provided a sufficient condition to guarantee the boundedness of  estimation error by each agent. The feasibility condition was analyzed through   an equivalent statement, which connects to 2s-sparse observability in the centralized framework in certain scenarios.

\bibliography{C:/work_at_kth/All_references}
\bibliographystyle{ieeetr}

\end{document}